\documentstyle[aps,eqsecnum,preprint,prd,epsfig,epsf]{revtex}
\begin{document}
\addtolength{\jot}{10pt}
\tighten
\def\bold#1{\setbox0=\hbox{$#1$}%
     \kern-.025em\copy0\kern-\wd0
     \kern.05em\%\baselineskip=18ptemptcopy0\kern-\wd0
     \kern-.025em\raise.0433em\box0 }
\def\slash#1{\setbox0=\hbox{$#1$}#1\hskip-\wd0\dimen0=5pt\advance
         to\wd0{\hss\sl/\/\hss}}
\newcommand{\spur}[1]{\not\! #1 \,}
\newcommand{\be}{\begin{equation}}
\newcommand{\ee}{\end{equation}}
\newcommand{\bea}{\begin{eqnarray}}
\newcommand{\eea}{\end{eqnarray}}
\newcommand{\nn}{\nonumber}
\newcommand{\dd}{\displaystyle}
\newcommand{\bra}[1]{\left\langle #1 \right|}
\newcommand{\ket}[1]{\left| #1 \right\rangle}
%
\draft
\preprint{\vbox{\hbox{BARI-TH/03-467 \hfill}
                \hbox{October 2003\hfill} }}
\vskip 2.0cm
\title{\bf
Nonfactorizable  contributions in $B$ decays to charmonium:\\
the case of $B^- \to K^- h_c$}
\author{P. Colangelo$^a$,  F. De Fazio$^a$ and T.N. Pham$^b$\\}
\vskip 1.0cm
\address{
$^a$ Istituto Nazionale di Fisica Nucleare, Sezione di Bari, Italy\\
$^b\;\;$ Centre de Physique Th\'eorique, \\
Centre National de la Recherche Scientifique, UMR 7644\\
\'Ecole Polytechnique, 91128 Palaiseau Cedex, France\\}

\maketitle
\vskip 0.5cm
\begin{abstract}
\noindent
Nonleptonic $B$ to charmonium decays
generally show deviations from the factorization predictions. 
For example,  the mode  $B^- \to K^- \chi_{c0}$ has been experimentally 
observed with sizeable branching fraction while its 
factorized amplitude vanishes. 
We investigate the role of rescattering effects mediated
by intermediate charmed meson production in this class of decay modes, and
consider $B^- \to K^- h_c$ with $h_c$ the 
$J^{PC}=1^{+-}$ $\bar c c$ meson.
Using an effective lagrangian describing interactions of pairs of
heavy-light $Q{\bar q}$ mesons with a quarkonium state, we relate this mode
to the analogous mode with $\chi_{c0}$ in the final state. 
We find ${\cal B}(B^- \to K^- h_c)$ large enough to be measured at the $B$ 
factories, so that this decay mode could be used to study 
the poorly known $h_c$. 
\end{abstract}
\vspace*{1cm}
\clearpage

\section{Introduction}\label{s:intr}
The precise test of the Standard Model description of CP violation 
in the $B$ sector is among the most challenging efforts pursued at
present experimental facilities. It goes through the measurement of many 
observables, such as CP asymmetries and $B$ meson branching fractions  which
are sensitive to CKM angles. In order to extract meaningful information 
from experimental data, a reduced theoretical uncertainty is required, and 
this is a particularly demanding task in the case of
nonleptonic $B$ decays for which a completely reliable
and general computational scheme has still to be developed.

For two-body nonleptonic $B$ decays, which concern us in the present paper,
the determination of the transition amplitude reduces to the calculation 
of the following matrix element of the effective Hamiltonian governing 
$B \to M_1 M_2$ \cite{Buras:1998rb}:
\be
A(B \to M_1 M_2) =
{G_F \over \sqrt 2} \sum_i \lambda_i c_i(\mu)
\langle M_1 M_2 | {\cal O}_i(\mu) | B \rangle  \,\,\,\ . \label{twobody}
\ee
In (\ref{twobody}) $\lambda_i$ are CKM matrix elements,
$c_i(\mu)$ Wilson coefficients evaluated at the scale $\mu$
and  ${\cal O}_i$  a set of four-quark operators.
So, neglecting  corrections to the r.h.s. of eq.(\ref{twobody}) that are
suppressed by inverse powers of $M_W$, the analysis of the decay
amplitude involves the calculation of hadronic matrix elements
of four-quark operators. 
The oldest prescription, which could be used to evaluate any generic form 
(\ref{twobody}), is the naive factorization ansatz that expresses
the matrix elements of four-quark operators 
as products of hadronic matrix elements of quark currents.

Let us consider $B^- \to K^- M_{\bar c c}$
which is pertinent to our discussion;  $M_{\bar c c}$
is a meson belonging to the charmonium system.
Neglecting the annihilation term which is suppressed by the CKM
factor $V_{ub}$,
the effective Hamiltonian $H_W$ driving the decay reads as
\be
H_W={G_F \over \sqrt 2} \Big\{V_{cb} V^*_{cs}
\Big( c_1(\mu) {\cal O}_1(\mu)+ c_2(\mu) {\cal O}_2(\mu) \Big)
-V_{tb} V^*_{ts} \sum_i c_i(\mu) {\cal O}_i(\mu) \Big \} + h.c.
\label{hamiltonian}
\ee
where
\bea
{\cal O}_1&=&(\bar c b)_{V-A} (\bar s c)_{V-A}  \nonumber \\
{\cal O}_2&=&(\bar s b)_{V-A} (\bar c c)_{V-A} \nn \\
{\cal O}_{3(5)}&=&(\bar s b)_{V-A} \sum_q(\bar q q)_{V-A [V+A]} \nn \\
{\cal O}_{4(6)}&=&(\bar s_i b_j)_{V-A} \sum_q(\bar q_j q_i)_{V-A [V+A]}
\label{effectiveoper} \\
{\cal O}_{7(9)}&=&{3\over 2}(\bar s b)_{V-A} \sum_q e_q (\bar q q)_{V+A [V-A]}
\nn \\
{\cal O}_{8(10)}&=&{3\over 2}(\bar s_i b_j)_{V-A}
\sum_q e_q (\bar q_j q_i)_{V+A[V-A]} \nonumber
\eea
($i$ and $j$ are color indices and
$(\bar q q)_{V\mp A}= \bar q \gamma^\mu (1 \mp \gamma_5) q$).
The corresponding expression of the factorized amplitude is:
\begin{equation}
{\cal A}_F(B^- \to K^- M_{\bar c c}) ={G_F \over \sqrt 2} V_{cb} V^*_{cs}
\Big[a_2(\mu)+\sum_{i=3,5,7,9} a_i(\mu) \Big]
\langle K^-|(\bar s b)_{V-A}|B^-\rangle
\langle M_{\bar c c} |(\bar c c)_{V \mp A}|0\rangle \label{matrixel}
\end{equation}
where  $a_i$ are combinations of Wilson coefficients:
$a_2=c_2+{c_1\over N_c}$ and $a_i=c_i+{c_{i+1}\over N_c}$, with $N_c$ 
the number of colors.

Eq.(\ref{matrixel}) shows the drawbacks of the
naive factorization approach:
first, the scale and scheme dependence of the Wilson coefficients
$c_i(\mu)$ is no longer compensated by a corresponding dependence of
the hadronic matrix element, and 
secondly, the product of hadronic matrix elements does not contain
any strong  phase.

Great amount of work has been done since this formulation of
factorization has been put forward, aiming either at finding alternative
procedures or at changing the ansatz itself. An improvement consists in
adopting a generalized factorization ansatz, with the Wilson coefficients
$a_i(\mu)$ replaced by effective (process independent) parameters 
$a_i^{\rm eff}$  to be fixed using experimental data.
In some cases this method reproduces the correct order of magnitude
of the branching ratios \cite{Neubert:1997uc}.
Other methods, such as QCD-improved
factorization \cite{Beneke:2000ry}, PQCD \cite{Keum:2000ph},
SCET \cite{Bauer:2001cu}, QCD sum rules \cite{shifman,Colangelo:2000dp},
can only be applied to selected classes of nonleptonic transitions.

In $B$ to charmonium decays, generalized factorization 
indicates the existence of sizeable nonfactorizable contributions. 
For example,
the experimental branching fraction  ${\cal B}(B \to K^- J/\psi)$
can be fitted using  $|a_2^{\rm eff}|=0.2-0.4$ depending
on the $B\to K$ transition form factor which parametrizes the matrix element
$\langle K^-|(\bar s b)_{V-A}|B^-\rangle$ in (\ref{matrixel})
\footnote{Since the other Wilson coefficients are numerically small,
one can safely consider only the contribution proportional to $a_2$.};
$|a_2^{\rm eff}|=0.38 \pm 0.05$ is obtained using the form factor 
in  \cite{Colangelo:1995jv}.
This must be compared to the value
$a_2=0.163 (0.126)$ computed for  ${\overline m}_b(m_b)=4.4$ GeV and
$\Lambda_{\overline{MS}}^{(5)}=290$ MeV in the
naive dimensional regularization (or 't Hooft-Veltman) scheme
\cite{Buras:1998rb}, a value which does not change significantly by varying
${\overline m}_b(m_b)$ and $\Lambda_{\overline{MS}}^{(5)}$.
The difference between  $a_2^{\rm eff}$ and  $a_2$ witnesses
the presence of nonfactorizable effects in this decay mode.

However, the most compelling evidence of deviation from factorization 
comes from the observation of  $B^- \to K^- \chi_{c0}$,
with $\chi_{c0}$ the lightest $\bar c c$ scalar meson. 
The measured branching fraction is:
\bea{\cal B}(B^- \to K^- \chi_{c0})&=&(6.0^{+2.1}_{-1.8}\pm 1.1)
\times 10^{-4} \;\;\;\;\;  \label{belledatum} \\
{\cal B}(B^- \to K^- \chi_{c0})&=&(2.4\pm 0.7) \times 10^{-4}
\label{babardatum} 
\eea
for BELLE \cite{Abe:2002mw} and BABAR \cite{Aubert:2002jn} Collaborations, 
respectively. While the experimental amplitude  evidently is
non-vanishing, the factorized amplitude  (\ref{matrixel}) is zero because
$\langle \chi_{c0} |(\bar c c)_{V \mp A}|0\rangle=0$. Interestingly,
the decay  occurs at a rate comparable to $B^- \to K^- J/\psi$ since, 
for example, 
${{\cal B}(B^- \to K^- \chi_{c0}) \over {\cal B}(B^- \to K^-
J/\psi)}=(0.60^{+0.21}_{-0.18}\pm 0.05\pm0.08)$
as reported by BELLE Collaboration \cite{Abe:2002mw}.

Analyses of the two modes $B^- \to K^- \chi_{c0}, K^- J/\psi$
in the framework
of QCD-improved factorization show that perturbative
QCD corrections are not able to reproduce the experimental branching
ratios, giving
either small contributions or producing infrared divergences,  a
signal of uncontrolled nonperturbative effects 
\cite{Cheng:2000kt}.  

In ref.\cite{Colangelo:2002mj} we investigated the possibility that
the deviation from the factorization predictions
in $B \to $ charmonium processes
may be ascribed to rescattering processes, essentially due to
intermediate charm meson exchanges  represented by
diagrams of the type depicted in fig.\ref{diagrams}.
Rescattering effects in heavy meson decays have been considered
recently, for example, to explain the observation of some
OZI-suppressed decays of  $\psi(3770)$  \cite{Achasov:vh}, or 
as  possible contributions to $B \to \pi \pi$ \cite{Kamal:ij},
$B \to K^{(*)} \pi$ \cite{Colangelo:1989gi},\cite{ciuchini}, 
$B_s \to \gamma \gamma$ \cite{Choudhury:1998rb}.
We found that rescattering effects could be sizeable, enough to produce
a large branching ratio as observed in $B^- \to K^- \chi_{c0}$.

Here we wish to reconsider the problem, since
other decays modes have vanishing factorized amplitude \cite{Diehl:2001xe}
and can be used to test the rescattering picture. One of them,  
$B^- \to K^- h_c$ with $h_c$ the lowest lying $J^{PC}=1^{+-}$ 
$\bar c c$ state, deserves particular attention.
The meson $h_c$ was searched \cite{Baglin:1986yd} and
observed \cite{Armstrong:1992ae} in $p{\bar p}$ annihilation,  
and  searched in $p-{\rm Li}$ interactions \cite{Antoniazzi:1993jz};
the  reported mass  and widths are
$m_{h_c} \simeq 3526 \,MeV$ and $\Gamma \le 1.1 \, MeV$. 
It is listed by the Particle Data Group among the particles requiring 
confirmation \cite{Hagiwara:fs}. If $B^- \to K^- h_c$  
proceeds with a sizeable rate, this decay could be used to study
the properties of  $h_c$  by looking  either at its 
hadronic transitions: $h_c \to J/\psi \pi^0$, $\rho^0 \pi^0$, 
$h_1 f_0(980)$,  $h_1 K {\bar K}$, \dots, or at its radiative 
decay modes: $h_c \to \eta_c \gamma$, $\chi_{c0} \gamma$, etc. 

This paper is devoted to such an investigation. Moreover, it 
aims at improving the analysis of rescattering effects in $B$ to 
charmonium transitions reducing the dependence of the rescattering
amplitude on unknown hadronic parameters, such as the strong couplings
among different mesons. We introduce 
an effective lagrangian describing the interaction of all the 
low-lying $\ell=1$ charmonium states to pairs of open charm $D^{(*)}_{(s)}$ 
mesons, based on the spin symmetry for the heavy quark in the infinite
heavy quark mass limit. This allows to express all the 
couplings in terms of  a single hadronic parameter, 
as shown in  Section \ref{s:efflag}. A similar relation is derived for 
the couplings of $\ell=0$ $\bar c c$ mesons to pairs of $D^{(*)}_{(s)}$.
Using such relations it is possible to analyze various rescattering amplitudes;
their calculation is reported in Sections \ref{s:resc} and \ref{s:calc},
while the conclusions concerning the possibility of using $B$ decays
to study the $h_c$ are drawn in the last Section.

\section{Model for charmed meson rescattering contributions}
\label{s:resc}
As for $B^- \to K^- \chi_{c0}$, the factorized amplitude 
$A_F(B^- \to K^- h_c)$ in (\ref{matrixel}) vanishes since the matrix element
$\langle h_c |(\bar c c)_{V \mp A}|0\rangle$ is zero due to 
conservation of parity and charge conjugation.  
This does not imply that the decay is forbidden, as other decay mechanisms
can be invoked, namely $h_c$ production {\it via} $\bar c c$ pair 
creation in the color octet configuration. 
From the hadronic point of view,
one can also consider the decay as proceeding by rescattering
processes induced by the same $(\bar b c)(\bar c s)$ effective
weak Hamiltonian in (\ref{hamiltonian}), processes that essentially account 
for a rearrangement of the quarks in the final state. Such effects are
not CKM suppressed, and their role must be assessed by explicit
(even though model dependent) calculations. Notice that color octet and
rescattering descriptions can represent two ways to describe the
same physics underlying the nonleptonic transition, looking from the
short-distance or the long-distance view points, respectively. 

We consider rescattering processes corresponding to the decay
chain $B^- \to X^0_{\bar u c} Y^-_{\bar c s} \to K^- M_{\bar c c}$, 
where $X$ and $Y$ are open charm resonances primarily produced in weak $B^-$
transitions. The lowest lying intermediate states $X^0_{\bar d c}$ and
$Y^-_{\bar c s}$ are the mesons $D_s^{(*)-}$ and $D^{(*)0}$, 
and we describe their rescattering by the exchange of
$D^{(*)}_{(s)}$ resonances, as depicted in fig.\ref{diagrams}. 

In order to analyze the diagrams in fig.\ref{diagrams} we need
the weak vertices $B \to D_s^{(*)} D^{(*)}$
and two strong vertices, one describing the coupling of a pair
of charmed mesons to kaon, the other one representing the
interaction of the charmonium state $h_c$ to a pair of $D_{(s)}^{(*)}$ mesons.
All nonperturbative quantities entering in such vertices can be related
to few hadronic parameters once the infinite heavy quark mass limit is 
adopted. 

In the following Section we analyze the couplings of the charmonium states
to pairs of open charm mesons. Here we consider
strong interactions of mesons $H_Q$ containing a single  heavy
quark $Q$ which can be described in the framework of the Heavy Quark
Effective Theory (HQET) \cite{hqet}, exploiting the heavy quark
spin and flavour symmetries holding in QCD for
$m_Q \to \infty$. In this limit the heavy quark four
velocity $v$ coincides with that of the hadron and it is
conserved by strong interactions \cite{Georgi:1990um}. Because of
the invariance under rotations of the heavy quark spin $s_Q$,
states differing only for the orientation of $s_Q$ are
degenerate in mass and form a doublet. When the
orbital angular momentum of the light degrees of freedom relative
to $Q$ is $\ell=0$, the two  states in the doublet have
spin-parity $J^P=(0^-,1^-)$ and correspond to $(D_{(s)},\,
D^*_{(s)})$,  $(B_{(s)},\, B^*_{(s)})$. This doublet can be
represented by a 4 $\times$ 4 matrix:
\be H_a=\left( {1 + \spur{v}
\over 2} \right) [M^\mu_a \gamma_\mu -M_a \gamma_5]  \,,
\label{hq}
\ee
with $M^\mu$ corresponding to the vector state and $M$
to the pseudoscalar one ($a$ is a light flavour index). The
fields $M_a$ and $M^*_a$ contain a factor $\sqrt{m_{M^{(*)}_a}}$, with 
$m$ the meson mass.

In the infinite heavy quark mass limit it is possible
to express weak as well as strong matrix
elements involving heavy mesons in terms of few universal quantities.
Let us consider the weak amplitude  $B^- \to D_s^{(*)-} D^{(*)0}$, for
which there is empirical evidence 
that the calculation  by factorization reproduces the main experimental
findings \cite{Luo:2001mc}.  Neglecting the contribution
of the operators ${\cal O}_{3-10}$ in (\ref{effectiveoper}) we can write:
\begin{equation}
\langle D_s^{(*)-} D^{(*)0} | H_W | B^- \rangle =
\displaystyle{G_F \over \sqrt{2}}V_{cb}V_{cs}^* a_1
\langle D^{(*)0} | (V-A)^\mu | B^- \rangle
\langle D_s^{(*)-}| (V-A)_\mu | 0 \rangle
\label{fact}
\end{equation}
with $a_1=c_1+c_2/N_c$.
In the infinite heavy quark mass limit,  the  matrix elements in (\ref{fact})
can be written in terms of a single form factor, the Isgur-Wise
function $\xi$, and  a single leptonic constant $\hat F$ \cite{hqet}.
The $B^- \to D^{(*)0}$ matrix elements read:
\begin{eqnarray}
<D^0(v^\prime)|V^\mu|B^-(v)>&=&\sqrt{m_B m_D} \; \xi(v \cdot v^\prime)
(v+v^\prime)^\mu\nonumber \\
<D^{*0}(v^\prime,\epsilon)|V^\mu|B^-(v)>&=& - i
\sqrt{m_B m_{D^*}} \; \xi(v \cdot v^\prime) \; \epsilon^*_{\beta} \;
\varepsilon^{\alpha \beta \gamma \mu} v_\alpha v^\prime_\gamma \label{BD} \\
<D^{*0}(v^\prime,\epsilon)|A^\mu|B^-(v)>&=&
\sqrt{m_B m_{D^*}} \; \xi(v \cdot v^\prime) \; \epsilon^*_{\beta}
[(1+v \cdot v^\prime) g^{\beta \mu}- v^\beta v^{\prime\mu}] \,\,\, ,\nonumber
\end{eqnarray}
$v$ and $v^\prime$ being $B^-$ and $D^{(*)0}$ four-velocities,
respectively,
$\epsilon$ the $D^*$ polarization vector and
$\xi(v \cdot v^\prime)$ the Isgur-Wise form factor.
The weak current for the transition from a heavy to a light quark
$Q \to q_a$, given at the quark level by
$\bar q_a \gamma^\mu(1-\gamma_5) Q$, can be written in terms of a heavy meson
and light pseudoscalars.
The octet of the light pseudoscalar mesons is represented by
 $\displaystyle \xi=e^{i {\cal M} \over f}$, with
\begin{equation}
{\cal M}=
\left (\begin{array}{ccc}
\sqrt{\frac{1}{2}}\pi^0+\sqrt{\frac{1}{6}}\eta & \pi^+ & K^+\nonumber\\
\pi^- & -\sqrt{\frac{1}{2}}\pi^0+\sqrt{\frac{1}{6}}\eta & K^0\\
K^- & {\bar K}^0 &-\sqrt{\frac{2}{3}}\eta
\end{array}\right )
\end{equation}
and $f\simeq f_{\pi}=131 \; MeV$, and the effective heavy-to-light current,
written at the lowest order in the light meson derivatives, reads:
\begin{equation}
L^\mu_a =
{\hat F \over 2} Tr[\gamma^\mu (1- \gamma_5) H_b \xi^\dagger_{ba}]
\,\,\, .
\end{equation}
In this way the matrix elements
$<0|\bar q_a \gamma^\mu (1-\gamma_5) c|D_a^{(*)}(v)>$, defined as
\begin{eqnarray}
<0|\bar q_a \gamma^\mu \gamma_5 c|D_a(v)> &=&f_{D_a} m_{D_a} v^\mu \nonumber
\\
<0|\bar q_a \gamma^\mu c|D^*_a(v,\epsilon)> &=&f_{D^*_a} m_{D^*_a}
\epsilon^\mu
\end{eqnarray}
can be related to the single quantity $\hat F$
since $f_{D_a}=f_{D^*_a}={\hat F \over \sqrt{m_{D_a}}}$.

It is also possible to write down an
expression for the strong couplings involving heavy mesons and the kaon.
The $D_s^{(*)} D^{(*)} K$ couplings, in the soft $\vec p_K \to 0$ limit,
can be related to a single low energy parameter $g$, as it turns out
considering the effective QCD lagrangian describing the
strong interactions between the heavy $D^{(*)}_a D^{(*)}_b$ mesons and
the octet of the light pseudoscalar mesons \cite{hqet_chir}:
\begin{equation}
{\cal L}_I = i \;
g \; Tr[H_b \gamma_\mu \gamma_5 {A}^\mu_{ba} {\bar H}_a]  \label{L}
\end{equation}
with the operator $A$ given by
\begin{equation}
{A}_{\mu ba}=\frac{1}{2}\left(\xi^\dagger\partial_\mu \xi-\xi
\partial_\mu \xi^\dagger\right)_{ba} \label{chirallag}
\end{equation}
and ${\bar H}_a=\gamma^0 H_a^\dagger \gamma^0$.
This allows to relate the $D_s^{(*)} D^{(*)} K$ couplings,
defined through the matrix elements
\begin{eqnarray}
<\bar D^0(p) K^-(q)|D_s^{*-} (p+q,\epsilon)>&=&
g_{D_s^{*-} \bar D^0 K^-} \, \, (\epsilon \cdot q) \nonumber\\
<\bar D^{*0}(p,\eta) K^-(q)|D_s^{*-} (p+q,\epsilon)>&=& i \,\,
\epsilon^{\alpha \beta \mu \gamma} \, p_\alpha \, \epsilon_\beta
\, q_\mu \eta^*_\gamma \,\, g_{D_s^{*-} \bar D^{*0} K^-}  \,\,\, ,\label{gddk}
\end{eqnarray}
to the  coupling $g$:
\begin{eqnarray}
g_{D_s^{*-} \bar D^{0} K^-}&=& 2 \sqrt{m_D m_{D_s^*}}
{\displaystyle g \over f_K} \nn \\
g_{D_s^{*-} \bar D^{^*0} K^-}&=& - 2 \sqrt{m_{D_s^*} \over m_{D^*}}
{\displaystyle g  \over f_K} \,\,\,\ .
\label{gcoupl}
\end{eqnarray}
All the above expressions are valid in the infinite limit
for the charm quark mass. We neglect corrections due to the
finite mass of the charm quark.

\section{Couplings of pairs of Heavy-Light Mesons to Quarkonium States}
\label{s:efflag}

The other strong vertex in the diagrams in fig.\ref{diagrams}
involves $h_c$ and a pair of open charm mesons. Also in this case
we exploit the infinite heavy quark mass limit.
For mesons with two heavy quarks $Q_1 {\bar Q}_2$
heavy quark flavour symmetry does not hold any longer, but
degeneracy is expected under rotations of the two
heavy quark spins. This allows us to build up heavy meson
multiplets for each value of the relative angular momentum $\ell$.
For $\ell=0$ one has a doublet comprehensive  of a pseudoscalar 
and a vector state, $\eta_c$ and $J/\psi$ in
case of charmonium. The corresponding 4 $\times$ 4 matrix reads
as \cite{Jenkins:1992nb}:
\be
R^{(Q_1 \bar Q_2)}=\left( {1 + \spur{v} \over 2} \right)[L^\mu \gamma_\mu -L
\gamma_5] \left( {1 - \spur{v} \over 2} \right) \label{rqq} \,,
\ee  
with $L^\mu=J/\psi$ and $L=\eta_c$ in case of $\bar c c$.
For $\ell=1$, four states can be built which are degenerate
in the heavy quark limit. The corresponding spin multiplet reads:
\be
P^{(Q_1 \bar Q_2)\mu}=\left( {1 + \spur{v} \over 2} \right) \left(
\chi_2^{\mu \alpha}\gamma_\alpha +{1 \over \sqrt{2}}\epsilon^{\mu
\alpha \beta \gamma} v_\alpha \gamma_\beta \chi_{1 \gamma}+ {1 \over
\sqrt{3}}(\gamma^\mu-v^\mu) \chi_0 +h_1^\mu \gamma_5 \right)\left( {1
- \spur{v} \over 2} \right) \label{pwave} 
\ee
where, in the case of $\bar c c$,
$\chi_2=\chi_{c2}$, $\chi_1=\chi_{c1}$ and $\chi_0=\chi_{c0}$
correspond the spin triplet, while the spin singlet is
$h_1=h_c$ \cite{Casalbuoni:1992fd}. Also the fields in (\ref{rqq}),
(\ref{pwave}) contain a factor $\sqrt m$, with $m$ the meson mass.

Using (\ref{rqq}) and (\ref{pwave}), together
with (\ref{hq}) representing the  heavy-light $Q_1 \bar q_a$ 
pseudoscalar and vector states, it is possible to write down 
the expressions for the effective couplings
between heavy-heavy mesons and pairs of heavy-light mesons 
we are interested in. 
For $\ell=1$ $Q_1 {\bar Q}_2$ state, the most general lagrangian 
describing the coupling to two heavy-light
mesons $Q_1 {\bar q_a}$ and $q_a{\bar Q}_2$ can be written as follows:
\be 
{\cal L}_1=i {\tilde g_1\over 2}
Tr \left[P^{(Q_1 \bar Q_2)\mu} {\bar H}_{2a} (\Omega_1 \gamma_\mu
+\Omega_2 {\mathrm v}_\mu) {\bar H}_{1a} \right] +h.c. + (Q_1
\leftrightarrow Q_2) \label{lagr1}  
\ee 
where $\Omega_1$ and $\Omega_2$ are two coefficients, 
$H_{1a}$ is given in  (\ref{hq}) and
$H_{2a}$ is the matrix describing the heavy-light mesons with quark content
$q_a{\bar Q}_2$: 
\be
H_{2a}= [M^{\prime \mu}_a \gamma_\mu - M^\prime_a \gamma_5] 
\left( {1 - \spur{v}\over 2} \right) \,\,\ .
\label{h2}
\ee
Due to the property $P^\mu v_\mu=0$ only the term proportional to $\Omega_1$
contributes, and therefore: 
\be 
{\cal L}_1= i {g_1 \over 2} Tr \left[P^{(Q_1 \bar Q_2)\mu}
{\bar H}_{2a} \gamma_\mu {\bar H}_{1a} \right] + h.c. + (Q_1
\leftrightarrow Q_2) \label{lagr1new} \,, 
\ee 
where $g_1={\tilde g}_1 \cdot \Omega_1$.
This expression accounts for the fact that the two heavy-light mesons
are coupled to the heavy-heavy state in S-wave, and therefore the  
matrix elements do not depend on their relative momentum. Moreover, 
this expression is invariant under independent rotations of the spin 
of the heavy quarks, representing the decoupling of the spin
in the infinite heavy quark mass limit. This can be easily seen considering
that under independent heavy quark spin 
rotations $S_1 \in SU(2)_{Q_1}$ and  $S_2 \in SU(2)_{Q_2}$
the following transformation properties hold for the various multiplets: 
\bea
H_{1a} \to S_1 H_{1a} & \hskip 0.5 cm &
{\overline H}_{1a} \to  {\overline H}_{1a} S_1^\dagger \,\, \nonumber \\
H_{2a} \to H_{2a}S_2^\dagger & \hskip 0.5 cm &
{\overline H}_{2a} \to  S_2{\overline H}_{2a}\,\, \nonumber \\
P^{(Q_1 \bar Q_2)\mu} \to  S_1 P^{(Q_1 \bar Q_2)\mu} &\hskip 0.5 cm & 
P^{(Q_1 \bar Q_2)\mu} \to P^{(Q_1 \bar Q_2)\mu} S_2^\dagger \label{su2spin} \\
R^{(Q_1 \bar Q_2)} \to  S_1 R^{(Q_1 \bar Q_2)}&\hskip 0.5 cm & 
R^{(Q_1 \bar Q_2)} \to R^{(Q_1 \bar Q_2)} S_2^\dagger \,\,\, .\nonumber  
\eea
Eq.(\ref{lagr1new}) shows that a unique coupling
describes the $P^\mu HH$ interaction, i.e. the same coupling controls
the interaction of heavy-light mesons both with the three $\chi_{c}$ states,
both with $h_c$. In particular,  from (\ref{lagr1new}) it follows that:
\bea 
<D^*_{(s)}(p_1,\epsilon_1)D_{(s)}(p_2)
|h_c(p,\epsilon)>&=&g_{D^*_{(s)}D_{(s)}h_c}(\epsilon^*_1 \cdot \epsilon) 
\nonumber \\
<D^*_{(s)}(p_1,\epsilon_1)D^*_{(s)}(p_2,\epsilon_2) |h_c(p,\epsilon)>
&=& i \, g_{D^*_{(s)}D^*_{(s)} h_c}\epsilon_{\alpha \beta \tau
\sigma} p^\alpha \epsilon^\beta \epsilon_1^{* \tau}
\epsilon_2^{*\sigma} \label{matrix1}
\eea
with
\bea 
g_{D^*_{(s)}D_{(s)}h_c}&=&- 2 g_1 \sqrt{m_{h_c} m_{D_{(s)}} m_{D^*_{(s)}}} 
\nonumber \\
g_{D^*_{(s)}D_{(s)}^*h_c}&=&  2 g_1
\sqrt{m_{D^*_{(s)}}^2 \over m_{h_c}} \;\;\; .\label{gddhc} 
\eea
Analogously:
 \bea
<D^*_{(s)}(p_1,\epsilon_1)D^*_{(s)}(p_2,\epsilon_2) |\chi_{c0}(p)>
&=& -g_{D^*_{(s)}D^*_{(s)} \chi_{c0}}(\epsilon_1^* \cdot
\epsilon_2^*) \nonumber \\ 
<D_{(s)}(p_1)D_{(s)}(p_2) | \chi_{c0}(p)>&=&
 -g_{D_{(s)} D_{(s)} \chi_{c0}} \label{matrix2} 
\eea
with
\bea
g_{D^*_{(s)}D^*_{(s)} \chi_{c0}}&=& - {2
\over \sqrt{3}} g_1 \sqrt{m_{\chi_{c0}}} m_{D^*_{(s)}}\nonumber \\
g_{D_{(s)}D_{(s)} \chi_{c0}}&=&-2 \sqrt{3}
g_1 \sqrt{m_{\chi_{c0}}} m_{D_{(s)}} \,\,\,\ .\label{gddchi} \eea 
The subscripts (1) and (2) refer to the meson with a charm 
and an anticharm quark, respectively; $\epsilon$,
$\epsilon_1$ and $\epsilon_2$ are polarization vectors.

Eqs.(\ref{matrix1})-(\ref{matrix2}) show that spin symmetry produces
stringent relations between the couplings of $\chi_{c0}$ and $h_c$
to open charm mesons, relations that we exploit below. Moreover, they also 
imply that the couplings of a single charmonium state to open  charm
pseudoscalar and  vector mesons are related
in absolute value and in sign as well, a property that allows a proper
analysis of the amplitudes in fig.1 where the relative signs between
different amplitudes  play an important role.

For the $\ell=0$ states represented by  the multiplet (\ref{rqq}), 
the interactions with the heavy-light
vector and pseudoscalar mesons proceed in P-wave and can be
described by a lagrangian containing a derivative term: 
\be
{\cal L}_2= {g_2\over 2} 
Tr \left[R^{(Q_1 \bar Q_2)} {\bar H}_{2a} \stackrel{\leftrightarrow}{{\spur
\partial}} {\bar H}_{1a} \right] +h.c. + (Q_1 \leftrightarrow Q_2)
\label{lagr2}  
\ee
which is also invariant under independent heavy quark spin rotations.
The action of the derivative produces a factor of  the residual
momentum $k$, i.e. the quantity for which the hadron and the heavy
quark four momentum differ: $M_H v_\mu=m_Q v_\mu +k_\mu$, 
$k$ being finite in the heavy quark limit.
The couplings of heavy-light charmed mesons to $J/\psi$ follow
from (\ref{lagr2}): 
\bea
<D^*_{(s)}(p_1,\epsilon_1)D^*_{(s)}(p_2,\epsilon_2)
|J/\psi(p,\epsilon)> &=& g_{D^*_{(s)}D^*_{(s)}\psi} \nonumber \\
 \big[(\epsilon \cdot \epsilon_2^*)(\epsilon_1^* \cdot q) &-&
 (\epsilon \cdot q)(\epsilon_1^* \cdot \epsilon_2^*) + (\epsilon \cdot
\epsilon_1^*)(\epsilon_2^* \cdot q) \big] \nonumber \\
<D_{(s)}^*(p_1,\epsilon_1)D_{(s)}(p_2) |J/\psi(p,\epsilon)>&=&
g_{D^*_{(s)}D_{(s)}\psi} \, i \, \epsilon_{\beta \mu \alpha \tau}
v^\beta \epsilon^\mu \epsilon_1^{* \alpha} q^\tau \label{matrix3}  \\ 
<D_{(s)}(p_1)D_{(s)}(p_2) |J/\psi(p, \epsilon)> &=&
g_{D_{(s)}D_{(s)}\psi} (\epsilon \cdot q) \nonumber
\eea 
where $q$ is the difference in the residual
momenta  of the two heavy-light charmed mesons $q=k_1-k_2$. Since
$p_1=m_{D^{(*)}_{(s)}} v +k_1$ and $p_2=m_{D^{(*)}_{(s)}} v +k_2$, then 
$q=p_1-p_2$. The three couplings in (\ref{matrix3})
are related to the single parameter $g_2$:
\bea
g_{D^*_{(s)}D^*_{(s)}\psi} &=& - 2 g_2\sqrt{m_\psi} m_{D^*_{(s)}} \nonumber \\
g_{D^*_{(s)}D_{(s)}\psi}&=& 2 g_2 \sqrt{m_\psi m_{D_{(s)}} m_{D^*_{(s)}}}
\label{gddpsi} \\
g_{D_{(s)}D_{(s)}\psi} &=& 2 g_2 \sqrt{m_\psi} m_{D_{(s)}}\,\,\, . \nonumber
\eea 

In principle, the couplings $g_1$ and $g_2$ must be computed
by nonperturbative methods.  An estimate can be
obtained invoking vector meson dominance (VMD) arguments. For
example, one can consider the $D$-meson matrix element of the scalar
$\bar c c$ current: $\displaystyle \langle D(v^\prime) | \bar c c
| D(v) \rangle$, assuming the dominance in the $t$-channel
of the nearest resonance, i.e. the scalar $\bar c c$ state,
and using the normalization of the Isgur-Wise form factor at the
zero-recoil point $v=v^\prime$. This allows  to express  $g_{D D
\chi_{c0}}$ in terms of the constant  $f_{\chi_{c0}}$ that
parameterizes the matrix element
\begin{equation}
\langle 0| \bar c c | \chi_{c0}(q)\rangle= f_{\chi_{c0}}
m_{\chi_{c0}} \,\,\,,  \label{fchi}
\end{equation}
obtaining:
\begin{equation}
g_{D D \chi_{c0}}= 2  {\displaystyle m_D m_{\chi_{c0}} \over
f_{\chi_{c0}}}  \,\,\,, \label{gchi}
\end{equation}
a relation which determines $g_1$ once $f_{\chi_{c0}}$ is known:
\be
g_1=-\sqrt {m_{\chi_{c0}}\over 3} {1\over f_{\chi_{c0}}} \,\,\, . \label{g1}
\ee
Adopting the same argument one can also obtain $g_2$ 
in terms of the $J/\psi$ leptonic constant $f_\psi$, defined by
$\langle 0| \bar c \gamma^\mu c | J/\psi(p,\epsilon)\rangle= f_\psi m_\psi
\epsilon^\mu$. From the VMD  result
\be
g_{DD\psi}={m_\psi \over f_\psi} \,\,\,  \label{ddpsiconst}
\ee
one gets:
\be
g_2={\sqrt {m_\psi} \over 2 m_D f_\psi} \,\,\, . \label{g2}
\ee

The input quantities for computing the diagrams in fig.\ref{diagrams}
are now available.
We have only to notice that the strong couplings described  above
do not account for the off-shell effect of the t-channel
$D^{(*)}_{(s)}$  particles, the virtuality of which 
can be large. As discussed in the literature, a method to account 
for such effect relies on the introduction of form factors:
\begin{equation}
g_i(t)=g_{i0}\,F_i(t)\,, \label{offshell}
\end{equation}
with $g_{i0}$ the corresponding on-shell couplings (\ref{gddk}),
(\ref{matrix1}) and
(\ref{matrix2}). A simple pole representation for $F_i(t)$:
\be
F_i(t)=\displaystyle{\Lambda_i^2 -m^2_{D^{(*)}} \over \Lambda_i^2-t}
\label{ff}\ee 
is consistent with QCD counting rules \cite{Gortchakov:1995im}.
We adopt it, keeping in mind that
the parameters in such form factors represent a source of
uncertainty in our results.

\section{Numerical analysis of $B^- \to K^- \lowercase{h_c}$}\label{s:calc}

\begin{figure}[ht]
\begin{center}
\mbox{\epsfig{file=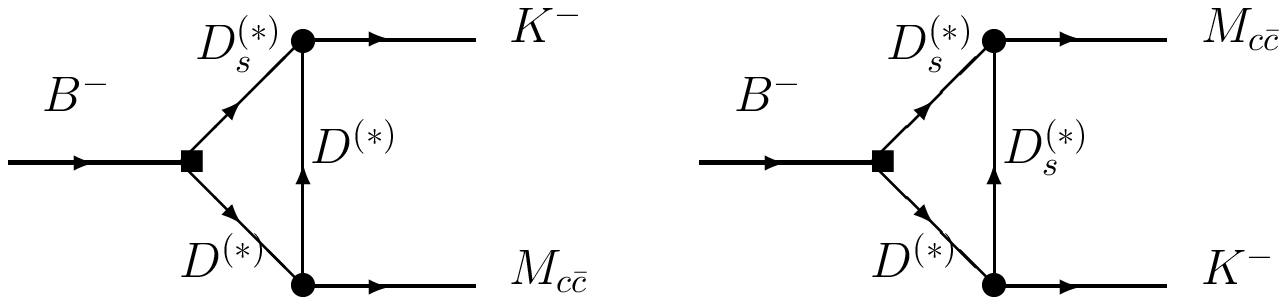, width=15cm}} \vspace*{-0.5cm}
\end{center}
\caption{Typical rescattering diagrams contributing to the
decay $B^- \to K^- M_{c{\bar c}}$, with $M_{c{\bar c}}$ a meson belonging
to the charmonium system. The boxes represent weak
vertices, the dots strong couplings.} \vspace*{1.0cm}
\label{diagrams}
\end{figure}

Considering the diagrams in fig.\ref{diagrams}  with $M_{\bar c c}=h_c$, 
there are ten
possible combinations of intermediate states corresponding to non-vanishing
strong vertices. Some of such diagrams vanish, since
the rescattering amplitude is parity
conserving and the final state $K^-h_c$ has positive parity
due to angular momentum conservation. As a consequence,
only the parity-violating weak decay amplitude contributes, hence only the
intermediate states $(D_s,D)$ and $(D^*_s,D^*)$ must 
be considered in fig.\ref{diagrams}.
The expression of the absorptive part of a generic diagram reads as:
\be
{\rm Im}{\cal A}=
{\sqrt{\lambda(m_B^2,m_{D_s^{(*)}}^2,m_{D^{(*)}}^2)}\over32 \pi m_B^2} 
\int_{-1}^{+1} dz {\cal A}(B^- \to D_s^{(*)-} D^{(*)0})
 {\cal A}(D_s^{(*)-} D^{(*)0}\to K^- h_c) \,\,\, .\label{amp}
\ee
In the case of the diagram in fig.\ref{diagrams} corresponding to 
$B^- \to D_s^- D^0 \to K^- h_c$ mediated by $D^{*0}$
the  expression (\ref{amp}) becomes:
\bea
{\rm Im}{\cal A}_1&=&{K \sqrt{m_B m_{D^0}} \over 32 \pi m_B^2}
\lambda^{1/2}(m_B^2,m_{D_{s}^-}^2,m_{D^0}^2)f_{D_{s}} \nonumber \\
&& (q \cdot\epsilon^*) \,\, 
\xi \left({m_B^2-m_{D_{s}}^2+m_{D^0}^2 \over 2 m_B m_{D^0}} \right)
\int_{-1}^1 dz {g_{D^* D_{s}K}(t) g_{D^* D h_c }(t) \over
t-m_{D^*}^2} f_1(z) \label{impart} \,\,\, , 
\eea 
with $K={G_F \over \sqrt{2}} V_{cb} V_{cs}^*a_1$,
$\lambda$ the triangular function, 
$q$  the kaon momentum and $\epsilon$ the $h_c$ polarization vector. 
The function $f_1$ is given by:
\bea
f_1(z)&=& -\left[ k^0 \left(1+{m_B \over m_D}
\right)-{m_{D_s}^2 \over m_D} \right] \big\{ \left( {m_K^2-q \cdot
k \over m_{D^*}^2}-1 \right) \nonumber \\ &-& {m_K^2-q \cdot k
\over m_{D^*}^2}{1 \over m_B |{\vec q}|^2} [(m_B
q^0-m_K^2)k^0-(m_B-q^0)q \cdot k] \big\}
\eea
with
$\displaystyle q^0={m_B^2+m_K^2-m_{h_c}^2 \over 2 m_B}$,  
$\displaystyle 
|{\vec q}|={\lambda^{1/2}(m_B^2,m_K^2,m_{h_c}^2) \over 2 m_B},$
$\displaystyle 
k^0 ={m_B^2+m_{D_{s}}^2-m_{D^0}^2 \over 2 m_B},$
$\displaystyle  
|{\vec k}|={\lambda^{1/2}(m_B^2,m_{D^0}^2,m_{D_{s}}^2) \over 2 m_B},$
$\displaystyle   q \cdot k = q^0 k^0 -|{\vec q}|~|{\vec k}|z$
and $\displaystyle t=m_K^2+m_{D_s}^2-2 q \cdot k$.
Expressions for the other diagrams can be worked out, analogously.
The t-dependence of the couplings is
given by eq.(\ref{ff}) with all $\Lambda_i$ put equal to a unique
parameter  $\Lambda$. 

We use $|V_{cb}|=0.042$ and $|V_{cs}|=0.974$,  the central
values reported by the Particle Data Group \cite{Hagiwara:fs},  and
$a_1=1.0$ as obtained from the analysis of exclusive $B \to D^{(*)}_s
D^{(*)}$ transitions  \cite{Luo:2001mc}.
Exploiting the heavy quark limit, we put
$f_{D^*_s}=f_{D_s}$ and use $f_{D_s}=240$ MeV \cite{Colangelo:2000dp}.
As for the Isgur-Wise form factor, the expression 
$\xi(y)= \left( {2 \over 1+y} \right)^2$ is compatible with the current 
results from the semileptonic $B \to D^{(*)}$ decays, and 
the product $V_{cb} \xi$ coincides with the experimental determination
reported in \cite{Luo:2001mc}. 

A comment is in order about the vertices $D^{(*)}_s D^{(*)}K$, 
expressed in terms of the coupling $g$ according to
(\ref{gcoupl}). An experimental determination of $g$
has been obtained by CLEO Collaboration measuring 
the full $D^*$ width and the $D^*$ branching fraction to $D \pi$.
The result is $g=0.59 \pm0.01 \pm 0.07$
\cite{Anastassov:2001cw}. Such a determination should be compared to
theoretical predictions ranging from $g\simeq 0.3$ up to $g \simeq 0.77$
\cite{gnew}.
Since the expressions of the rescattering amplitudes always contain
the product of $g$ and the form factor (\ref{ff}), we use  the
central value of $g$ obtained by experiment,
leaving to the parameter $\Lambda$
the task of spanning the range of possible variation of the coupling.

For $g_1$  we use eq. (\ref{g1})  together the QCD sum rule result 
$f_{\chi_{c0}}=510 \pm 40 $ MeV \cite{Colangelo:2002mj}. The coupling
$g_2$ can be obtained using (\ref{g2}) and the experimental
value $f_{J/\psi}=405 \pm 14 $ MeV.
The VMD determination of the $J/\psi$ couplings is reproduced by QCD sum rule 
and constituent quark model analyses \cite{Deandrea:2003pv}. Relating
the various couplings to  $g_1$ and $g_2$ we use $m_D=m_{D_s}$ and
$m_{D^*}=m_{D^*_s}$.

Eq. (\ref{amp}) allows us to compute the imaginary part of the
rescattering diagrams. The determination of the real part is more uncertain.
A dispersive integral may be used:
$\displaystyle
{\rm Re} {\cal A}_i (m_B^2) = {1 \over \pi} PV \int_{s_{th}^{(i)}}^{+\infty}
{ {\rm Im} {\cal A}_i (s^\prime) \over s^\prime - m_B^2} d s^\prime$
with the thresholds $s_{th}^{(i)}$ given by:
$s_{th}^{(i)}=(m_{D^{(*)}_s}+m_{D^{(*)}})^2$ for any specific diagram.
Assuming that the integrals are dominated by the region
close to the pole $m_B^2$,  so that they can be computed by using
a cutoff not far from the $B$ meson mass, we obtained 
for  $B^- \to K^- \chi_{c0}(J/\psi)$ that the real parts of the
amplitudes are approximately equal to the imaginary parts, with large
uncertainties due to the cut-off procedure
\cite{Colangelo:2002mj}. For this reason we account for
the real part of the amplitudes considering them as fractions 
of the imaginary part varying from $0$ to $100 \%$, i.e.
we include their contribution to the
final result considering the range from ${\rm Re} {\cal A}_i=0$ up to 
${\rm Re} {\cal A}_i \simeq  {\rm Im} {\cal A}_i$. Such an uncertainty
cannot be removed in our approach and will affect the final result.

A parameter is left in our analysis, i.e. the constant $\Lambda$
in the form factors (\ref{ff}).
One would expect $\Lambda$ of the order of the mass of
radial excitations of the charmed mesons. It is possible to constrain
the range of values for such a parameter considering
rescattering contributions to $B^- \to K^- J/\psi$, where the sum
${\cal A}(B^- \to K^- J/\psi)={\cal A}_{fact}+{\cal A}_{resc}$ is bounded 
by the experimental measurement of the branching
fraction ${\cal B}(B^- \to K^- J/\psi)$. If one considers the range
$2.6-3$ GeV for $\Lambda$ one gets a rescattering contribution not exceeding
the experimental bound.
Moreover, one can consider
$B^- \to K^- \chi_{c0}$ as provided only by rescattering effects, 
repeating the analysis in \cite{Colangelo:2002mj}, with the difference 
of using the relations (\ref{matrix2}) which
imply a factor of $3$ between the couplings of
$\chi_{c0}$ to pairs of $D$ and $D^*$ mesons, dictated by the spin symmetry.
With this factor into account, one gets a branching fraction
compatible with the experimental result from BABAR  if the parameter $\Lambda$
is varied around $3.0$ GeV.

\begin{figure}[h]
\begin{center}
\vspace*{0.4cm}
\mbox{\epsfig{file=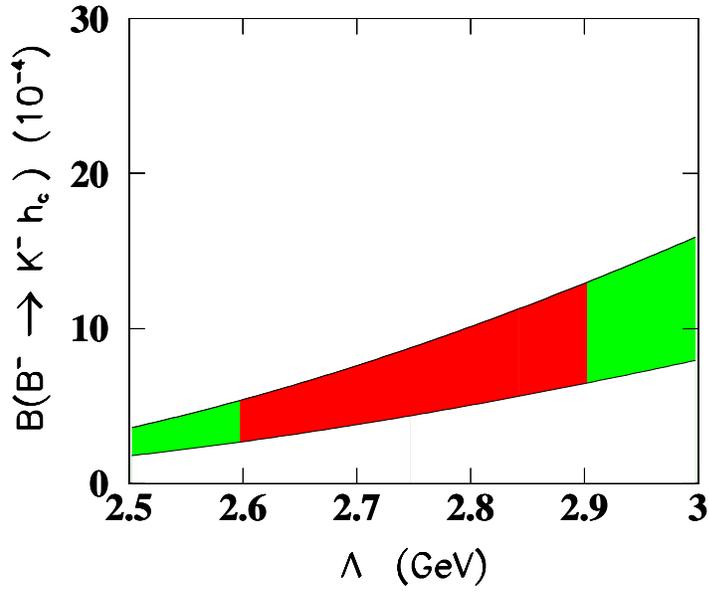, width=11cm}}
\end{center}
\caption{Branching fraction ${\cal B}(B^- \to K^- h_{c})$ versus 
the parameter $\Lambda$.
The lowest curve corresponds to ${\rm Re} {\cal A}_i=0$, the highest one to
${\rm Re} {\cal A}_i={\rm Im} {\cal A}_i$.
The dark region corresponds to the result (\ref{result-hc}).}
\label{br_hc}
\end{figure}

Provided with such constraints  we analyze  $B^- \to K^- h_c$. In
fig.\ref{br_hc} we plot the branching ratio  obtained considering
the rescattering amplitudes as a function of $\Lambda$.  We find a region
that can be represented by the interval:
\be
{\cal B}(B^- \to K^- h_c)=(2 - 12) \times 10^{-4} \,\,\, , 
\label{result-hc}
\ee
where the range of values accounts for the uncertainty on 
the dispersive part of the rescattering amplitudes and on the variation of the
parameter $\Lambda$. This result suggests that $B^- \to K^- h_c$ occurs with a
rate large enough to produce a signal at the B-factories, as 
discussed in the next Section. Moreover, the outcome (\ref{result-hc})
implies that $B^- \to K^- h_c$
represents a sizeable fraction of the inclusive $B^- \to X  h_c$ mode, 
the branching ratio of which, 
estimated considering the production of the $c{\bar c}$ pair in $h_c$
in the color-octet state, is:
${\cal B}(B^- \to  h_c X)=(13 - 34) \times 10^{-4}$
\cite{Beneke:1998ks}.

The theoretical uncertainties affecting our results are
related to the  poorly known values of some of the input 
parameters and to the basic assumptions adopted in the calculation.
While the numerical values of several  parameters (namely, the
strong couplings among heavy mesons) can be made more precise 
using new experimental or theoretical information, it is difficult to
assess the actual size of the uncertainties related 
to the computational scheme
we have used in  evaluating rescattering effects. The main 
uncertainty in the numerical results
is due to large cancellations between different amplitudes,
which individually turn out to be of similar size. 
This is common to
calculations involving hadronic degrees of freedom, and it is not easy
to envisage a procedure for reducing or controlling the final error.
Another uncertainty is due to the neglect, in the calculation of
diagrams in fig.\ref{diagrams}, of contributions of higher
resonances and of many-particle intermediate states,  even though a 
minor role can be presumed for  higher resonances 
since the corresponding
universal form factors and leptonic decay constants are expected to be smaller
than for low-lying states.

Bearing such uncertainties in mind  we can conclude
that rescattering terms may contribute to the 
nonfactorizable effects observed in $B \to $ charmonium transitions.

\section{Remarks about the observation of  $B^- \to K^- \lowercase{h_c}$ and
conclusions}\label{s:conc}
Let us discuss few phenomenological consequences 
of our study, coming  first to the possibility 
of detecting and studying $h_c$ using $B$ decays. 

As mentioned in the Introduction, observation of 
$h_c$ has been reported in $p{\bar p}$ annihilation and in
$p-{\rm Li}$ interactions, where the meson is produced through $q{\bar q}$
annihilation in three gluons. 
Other production mechanisms are possibile at $e^+-e^-$ machines, namely 
{\it via} $\psi^\prime$ intermediate production. For example, 
one can consider the radiative decay   
$\psi^\prime \to \eta^\prime_c \gamma$ with the subsequent transition
$\eta^\prime_c \to h_c \gamma$ as feasible to obtain a sample of $h_c$. 
Another possibility is the 
hadronic decay mode $\psi^\prime \to h_c \pi^0$. In this case
the estimated branching ratio is rather sizeable: 
${\cal B}(\psi^\prime \to h_c \pi^0) \simeq {\cal O}(10^{-3})$
\cite{Kuang:2002hz}, and therefore one could consider 
the investigation affordable, e.g., at a charm factory; however,
a low  $\pi^0$ reconstruction efficiency could severely limit 
the possibility of studying $h_c$ produced by this decay chain.

As for $h_c$ produced in $B$ decays, one could access the meson
looking either at its hadronic modes:
$h_c \to J/\psi \pi^0$, $\rho^0 \pi^0$, $h_1 f_0(980)$, $h_1
K {\bar K}$, \dots ,  or at its radiative modes: 
$h_c \to \eta_c \gamma$, $\chi_{c0} \gamma$, etc.  
In particular,  the channel $h_c \to \eta_c \gamma$ seems promising,
as noticed by Suzuki \cite{Suzuki:2002sq}.
Its branching ratio, estimated assuming
that the $h_c$ wave function close to the origin is the same 
as that of  $\chi_{c1}$, is large:
${\cal B}(h_c \to \eta_c \gamma) \simeq 0.50 \pm 0.11$ 
\cite{Suzuki:2002sq}.
A similar result:  ${\cal B}(h_c \to \eta_c \gamma) = 0.377$
\cite{Godfrey:2002rp} 
is obtained using the charmonium wave functions 
parameterized in ref.\cite{Godfrey:xj}. These two predictions, together with
the experimental datum for
${\cal B}(\eta_c \to K {\bar K} \pi)$, allow us to translate our
result  (\ref{result-hc}) in a prediction for the decay chain
$B^- \to K^- h_c \to K^- \eta_c \gamma \to K^- (K {\bar K} \pi) \gamma$
which can be studied at a $B$ factory:
\be
{\cal B}(B^- \to K^- h_c \to K^- \eta_c \gamma \to K^- (K {\bar K} \pi)
\gamma)= (4 - 26) \times 10^{-6} \,\, ,  \label{chain}
\ee
a result within the reach of current experiments. 
It is worth noticing that the investigation of
this particular decay chain is favoured by the rather accurate 
knowledge of the $\eta_c$ hadronic decays,
and by the fact that one could use the $\eta_c$ mass and the photon direction
to discriminate the signal from the background.

Coming to the role of rescattering effects in $B \to $ charmonium
transitions, we have found that they can be effective, and are able to
produce for the mode $B^- \to K^- h_c$ a branching fraction  
comparable with that of $B^- \to K^- \chi_{c0}$. Further
evidence for the presence of large nonfactorizable contributions
in B decays with charmonium in the final state can be obtained by
looking at other decay modes. One possibility is 
$B^- \to K^- \psi(3770)$ which, because of the smallness of the 
leptonic decay constant $f_{\psi(3770)}$, is predicted by
the factorization model with a tiny branching ratio.  
The observation of this decay mode with a sizeble branching fraction
${\cal B}(B^- \to K^- \psi(3770))=(0.48\pm0.11\pm0.12)\times 10^{-3}$
\cite{Abe:2003zv}
represents a further evidence of the presence of large 
nonfactorizable contributions. In our approach, using 
$g_{D D \psi(3770)}=14.94 \pm 0.86$ obtained from the width of 
$\psi(3770)$, we would get 
${\cal B}(B^- \to K^- \psi(3770))=(0.9-4)\times 10^{-4}$,
consistent with the experimental datum considering the large
theoretical uncertainty. Similar conclusion applies to
$B^- \to K^- \chi_{c2}$ with $\chi_{c2}$ the $J^{PC}=2^{++}$
state of the charmonium system, the amplitude of which also vanishes
in the factorization approach. The observation of this decay mode with 
branching fraction comparable to ${\cal B}(B^- \to K^- \chi_{c0})$ 
and ${\cal B}(B^- \to K^- h_c)$  would support  the rescattering picture.

\vspace*{2cm}
\noindent {\bf Acnowledgments.}
We acknowledge partial support from the EC Contract No.
HPRN-CT-2002-00311 (EURIDICE).

\newpage
\end{document}